\documentclass{aa}  

\usepackage{array}
\usepackage{booktabs}
\usepackage{graphicx}
\usepackage{txfonts}
\usepackage{color}
\usepackage{amsmath}	
\usepackage{mathtools}

\usepackage{hyperref}
\hypersetup{
    colorlinks=true,        
    linkcolor=blue,         
 citecolor=cyan,      
 filecolor=magenta,      
 urlcolor=magenta            
}
\def\gsim{\mathrel{\raise.5ex\hbox{$>$}\mkern-14mu
             \lower0.6ex\hbox{$\sim$}}}
\def\lsim{\mathrel{\raise.3ex\hbox{$<$}\mkern-14mu
             \lower0.6ex\hbox{$\sim$}}}
             
\newcommand{\defeq}{\vcentcolon=}
             
\begin{document}

	\title{Flares from plasmoids and current sheets around Sgr~A*} 
 \titlerunning{Sgr~A* flares}
 \authorrunning{Dimitropoulos et al.}
	
	\author{I. Dimitropoulos\thanks{up1098378@upatras.gr}
		\inst{1,2},
		A. Nathanail\thanks{anathanail@academyofathens.gr}
		\inst{2},
		M. Petropoulou
		\inst{3,4},
		I. Contopoulos\inst{2},
  C. M. Fromm\inst{5,6,7}
	}
	
	\institute{Department of Physics, University of Patras, Rio 26504, Greece  
		\and
		     Research Center for Astronomy and Applied Mathematics, Academy of Athens, Athens 11527, 
		     Greece
		\and
		     Department of Physics, National and Kapodistrian University of Athens, University Campus 
		     Zografos, GR 15784, Athens, Greece
       \and
        Institute of Accelerating Systems \& Applications, University Campus Zografos, GR 15784, Athens, Greece
         \and
         Institut f\"ur Theoretische Physik und Astrophysik,
  Universit\"at W\"urzburg, Emil-Fischer-Strasse 31, 97074 W\"urzburg,Germany 
  \and
        Institut f\"ur Theoretische Physik, Goethe Universit\"at
  Frankfurt, Max-von-Laue-Str.1, 60438 Frankfurt am Main, Germany 
  \and
        Max-Planck-Institut f\"ur Radioastronomie, Auf dem H\"ugel 69,D-53121 Bonn, Germany }
	
	
	\abstract
	{The supermassive black hole Sgr~A* at the center of our galaxy produces repeating near-infrared flares that are observed  by ground and space based instruments. This activity has been simulated in the past with Magnetically Arrested Disk (MAD) models which include stable jet formations. The present study uses a different approach in that it considers a Standard and Normal Evolution (SANE) multi-loop model that lacks a stable 
    jet structure.}
	{The main objective of this research is to identify regions that contain current sheets and high magnetic turbulence, and to subsequently generate a 2.2~$\mu$m light curve generated from non-thermal particles. These aims require the identification of areas which contain current sheets and high magnetic turbulence, and the averaging of the magnetization in the regions surrounding these areas. Subsequently, Particle-in-Cell (PIC) fitting formulas are applied to determine the non-thermal particle distribution and to obtain the sought after light curve. Additionally, we investigate the properties of the flares, in particular, their evolution during flare events, and the similarity of flare characteristics between the generated and observed light curves.
 }
	{2D GRMHD simulation data from a SANE multi-loop model is employed, and thermal radiation is introduced to generate a 230 GHz light curve. Physical variables are calibrated to align with the 230~GHz observations. Current sheets are identified by analyzing toroidal currents, magnetization, plasma $\beta$, density, and dimensionless temperatures. The evolution of current sheets during flare events is studied, and higher-energy non-thermal light curves are calculated, focusing on the 2.2~$\mu$m  near-infrared range.}
	{We obtain promising $2.2~\mu$m lightcurves whose flare duration and spectral index behavior align well with observations. Our findings support  the association of flares with particle acceleration and
nonthermal emission in current sheet plasmoid chains and in the boundary of the disk inside the funnel above and below the central black hole.  }
	{}
	
    \maketitle
    \lhead{Dimitropoulos et al.}
    \rhead{Sgr~A* flares}
\section{Introduction}
	In the center of the Milky Way at a distance of $D\simeq8.3$~kpc lies Sagittarius A* (Sgr~A*), a $4.2\times10^6~M_\odot$ supermassive black hole (hereafter SMBH; 
 \citealt{abuter2022mass}). The accretion disk
around it is optically thin, with collisionless high-temperature low-density plasma. Its bolometric luminosity is in the sub-Eddington range of
$L\sim 10^{36}~{\rm erg\ s}^{-1}$ (\citealt{genzel2010galactic}). The emission from Sgr~A* is
systematically monitored over a very wide range of frequencies, allowing
us to investigate the dynamics of the disk in the
vicinity of a SMBH. 

Observations from the GRAVITY Collaboration have shown a hot spot
rotating around Sgr~A* with a period of $\sim60$~minutes
 at a distance of a few gravitational radii (hereafter $r_g$) from the black
hole (\citealt{abuter2018detection}). Sgr~A* is known to exhibit variability in the near infrared
which manifests itself in the form of several flares over a single day (\citealt{ghez2005first};
\citealt{do2019unprecedented}). For the first time, GRAVITY observations
revealed that a flare in Sgr~A* coincided with a hot spot moving around
the central black hole. Several investigations suggest that flares
originate in highly magnetized structures that are formed in the
innermost region of the black hole and produce synchrotron radiation
(\citealt{dodds2009evidence}; \citealt{boehle2016improved};
\citealt{ponti2017powerful}; \citealt{chatterjee2021general};
\citealt{scepi2022sgr}). 

Analytical models have been applied to study the trajectory of the hot
spots in the resulting lightcurve and showed the importance of the effect
of gravitational lensing (\citealt{broderick2005imaging};
\citealt{younsi2015variations}; \citealt{baubock2020modeling};
\citealt{ball2021}). As shown by \citet{matsumoto2020origin}, assuming that its motion is in the equatorial plane, the GRAVITY hot spot follows a circular path at
super-Keplerian velocity. However, an outflowing conical sub-Keplerian orbit can fit these observations equally well \citep{Antonopoulou2024}.

Lately,  the theoretical astrophysics community has increasingly turned
to numerical simulations, specifically to general relativistic
magnetohydrodynamic simulations (hereafter GRMHD), in order to gain a comprehensive
understanding of the dynamics involved in accreting black holes. Observations of
flares and moving hot spots further emphasizes the significance of
considering both magnetic fields and fluid dynamics within the extreme
gravitational environment near a black hole (\citealt{dexter2020sgr};
\citealt{porth2021flares}; \citealt{chatterjee2021general};
\citealt{cemeljic}; \citealt{scepi2022sgr};
\citealt{mellah2023reconnection}; \citealt{Lin2024}).

Numerical investigations explore the formation of current sheets in the 
vicinity of a black hole and the subsequent production of plasmoids and 
plasmoid chains (\citealt{nathanail2020};
\citealt{nathanail2022_flares}; \citealt{ripperda2022black}). Current sheets are responsible for particle acceleration and
the generation of variable non-thermal radiation.  
These models need to be supported by numerical studies of
turbulence and magnetic reconnection in collisionless plasmas that self-consistently capture the dynamical interplay between particles and fields on the kinetic plasma scales. Such investigation can be performed with Particle-in-Cell (PIC) simulations
(\citealt{drake2012power}; \citealt{guo2014formation}; \citeyear{guo2015particle}; \citealt{sironi2014relativistic}; \citealt{shay2014electron}; 
\citealt{dahlin2014mechanisms}; \citealt{li2015nonthermally}; \citeyear{Li2023}; \citealt{petropoulou2018steady};
\citealt{werner2018non}; \citealt{ball2018slope}; \citealt{comisso2019interplay}).
In particular, these investigations help us study the acceleration processes in detail by tracking individual particles, but also characterize the properties of the particle distribution function, e.g. shape of distribution (power law, log-parabolic), fraction of energy carried by non-thermal particles, maximum particle energy and others 
(e.g., \citealp{werner2018non, ball2018slope,petropoulou2019,zhang2023}). Some of the particle properties, like the slope of the power-law distribution, can be related, through empirical expressions, to the  local
plasma properties, such as magnetization $\sigma \coloneqq B^2/\rho$ ( where $B$ is the magnetic field strength and $\rho$ is the density) and plasma~$\beta \coloneqq 2P/B^2$ (where $P$ is the pressure).
These prescriptions can then be
combined with GRMHD simulations to anticipate the generation of non-thermal, 
high-energy electron distributions with properties depending on the local physical conditions (\citealt{chatterjee2021general}; \citealt{scepi2022sgr}; 
\citealt{aimar2023_plasmoids}; \citealt{lin2023_plasmoids};
\citealt{mellah2023reconnection}).
The way to connect  GRMHD simulations and PIC results is  
 the following: after the plasma accretion process has reached an 
 inflow equilibrium \citep{dexter2020sgr}, the thermal radiation is
 attributed to the hot electrons with a parameterized temperature derived from the simulation ion temperature \citep{moscibrodzka2016general}. In order to include non-thermal electrons to radiate, formulas from PIC simulations are employed.
Non-thermal electron distributions are assumed in each cell  depending on magnetization (for $\sigma <5$) and plasma $\beta$ \citep{Davelaar2019, Fromm2022, Cruz-Osorio2022}. This study proposes a 
novel approach for incorporating non-thermal particles into simulation results. Specifically, current sheets are first identified (see also \citealt{Vos202}), followed by 
characterizing their environment in terms of averaged quantities of magnetization $\sigma$  and plasma~$\beta$. Finally, non-thermal particles are sourced 
from the plasma within the current sheet.
In this paper, we utilize the 2D GRMHD simulations of
\citet{nathanail2020} to investigate the ability of the current sheet and
plasmoids to produce the observed non-thermal flaring activity of Sgr~A*. 
A radiation model is applied to produce lightcurves at 230~GHz
(thermal radiation from the disk) and 2.2~$\mu$m (non-thermal radiation
from current sheets; see also \citealt{scepi2022sgr}). The structure of the 
paper is as follows: Section~2 describes the methodology of our investigation divided in three subsections in which we introduce 
the GRMHD simulation setup (2.1), the modeling of the radiation (2.2), and the 
method with which we determine current sheets and their environment from the GRMHD simulation (2.3). In Section~3 we present our results for the thermal (3.1) and non-thermal radiation (3.2). Finally,  in Section~4 we present our conclusions and discuss  prospects for future work.

\section{Numerical setup}
	
	The numerical methods employed for the GRMHD simulation in this study
	closely mirror those used in previous research (\citealt{nathanail2020}).
	We initialize the simulation with magnetic field configurations designed to
	generate multiple current sheets as the accretion system evolves. The
	formation of current sheets occurs in the vicinity of the black hole, and
	their subsequent reconnection gives rise to the development of multiple
	plasmoids and plasmoid chains. In order to assess the radiation characteristics of the current sheets formed and the subsequent generation of plasmoids, we employ a radiation proxy model. 	This model enables the calculation of both thermal and non-thermal radiation at 230~GHz and $2.2~\mu$m, respectively.
	
\subsection{GRMHD simulation}
	
	\label{sec:GRMHD}
	
	The accretion disk surrounding a black hole can be conceptualized as a
	hydrodynamic system in the context of curved spacetime, where the
	magnetic field significantly influences its dynamics. A GRMHD code
	integrates these physical properties into a unified numerical simulation
	and allows a robust evolution in time where the accretion and accumulation
	of magnetic flux produce energetic phenomena. We employ  the BHAC code (\citealt{porth2017}) which solves
	the ideal MHD equations in general relativity, namely
		\begin{equation}\label{grmhd1}	
			\nabla_{\mu}(\rho u^{\mu}) = 0
	\end{equation}
		\begin{equation}\label{grmhd2}
			\nabla_{\mu}T^{\mu \nu} = 0
	\end{equation}
		\begin{equation}\label{grmhd3}
			\nabla_{\mu} ~^*\!F^{\mu \nu}= 0
	\end{equation}
	where $T^{\mu \nu}$ and $~^*\!F^{\mu \nu}$ are the energy momentum tensor
	and the dual of the Faraday tensor, respectively. Here, we denote with $\rho$
	the rest-mass density and with $u^{\mu}$ the fluid four-velocity. Some
	important characteristics of the code are the following: it implements
	second-order high-resolution shock-capturing finite-volume methods and
	adaptive mesh-refinement (AMR) wherever needed. It also implements
	constrained-transport  (\citealt{del2007echo}) to guarantee a divergence-free
	magnetic field (\citealt{olivares2019constrained}).
	
	Furthermore, the model is axisymmetric (2D), and the coordinates of the
	initial torus-like plasma distribution  are spherical ($r \times \theta
	\times \phi$). We used a logarithmic radial grid and the domain extends
	out to $2500~r_g$. The resolution of the model is $4096 \times 2048 \times 1$ cells. The torus at the initial equilibrium state has constant
	specific angular momentum  $l= 4.28$ (\citealt{fishbone1976relativistic}).
	The inner radius was set to $r_{in} = 6~r_g$ and the pressure maximum
	radius to $r_{max} = 12~r_g$. All quantities are calculated in
	geometrized units ($G = c = 1$) in which the gravitational radius is
	equal to $r_g=M$. We considered a Kerr black hole with
	dimensionless spin $a=J/M^2=0.93$, where $J$ and $M$ are the angular momentum
	and mass of the black hole respectively. The radius of the event horizon is
	$r_h=1.341~r_g$, and there are 29 grid cells inside the horizon.
	
\begin{figure}
		\centering
		\includegraphics[width=0.6\columnwidth]{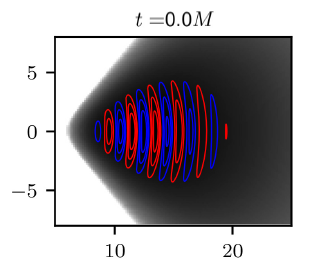}
		\caption{Initial magnetic field (lines) and density (color) configuration in the initial torus. Axis coordinates calculated in $r_g$. Red and blue colors denote clockwise and anticlockwise direction of the magnetic field. }
		\label{fig:magnetic_initial}
	\end{figure}

	The initial conditions for the
	magnetic field  inside the torus consists of several loops of interchangeable clockwise-anticlockwise orientation between neighboring loops (see Fig.~\ref{fig:magnetic_initial}).    In order to generate a magnetic field
	topology as described above we used a vector potential $A_{\phi}$:
		\begin{equation}\label{potential1}
			A_\phi \propto A \times B\ ,
	\end{equation}
 with
		\begin{equation}\label{potential2}	
			A\equiv {\rm max}(\rho / \rho_{max} - 0.2,0)\ ,
	\end{equation}
		\begin{equation}\label{potential3}
			B \equiv \cos{[(N-1)\theta] \sin{[2\pi(r-r_{in})/\lambda_r]}}\ ,
	\end{equation}
	where $N$ is the number of poloidal loops and $\lambda_r$ their
	characteristic length scale, and $\rho_{max}$ is the maximum rest mass
	density of the disk.
	
	The above model is characterized as SANE multi-loop and has  important
	differences from a classic Magnetically Arrested Disk (MAD) model. In a
	classic MAD there is a steady jet, whereas in the SANE multi-loop model no steady polar outflow develops. We have not yet observed a jet from Sgr~A* and it is an open
	question in which direction our models should go for the interpretation
	of phenomena near the black hole. As discussed in
	\citet{nathanail2022magnetic}, images at $43$ and $86$ GHz will produce an extended stable jet base in the case of
	MAD model, whereas in the case of the
	SANE multi-loop model, the images will possibly have an extended jet-like structure only in the 	flaring state.  We expect that new images of the Sgr~A* black hole will further clarify the landscape of pertinent models. 
	
   \begin{figure*}[h!]
        \centering
        \includegraphics[width=1.3\columnwidth]{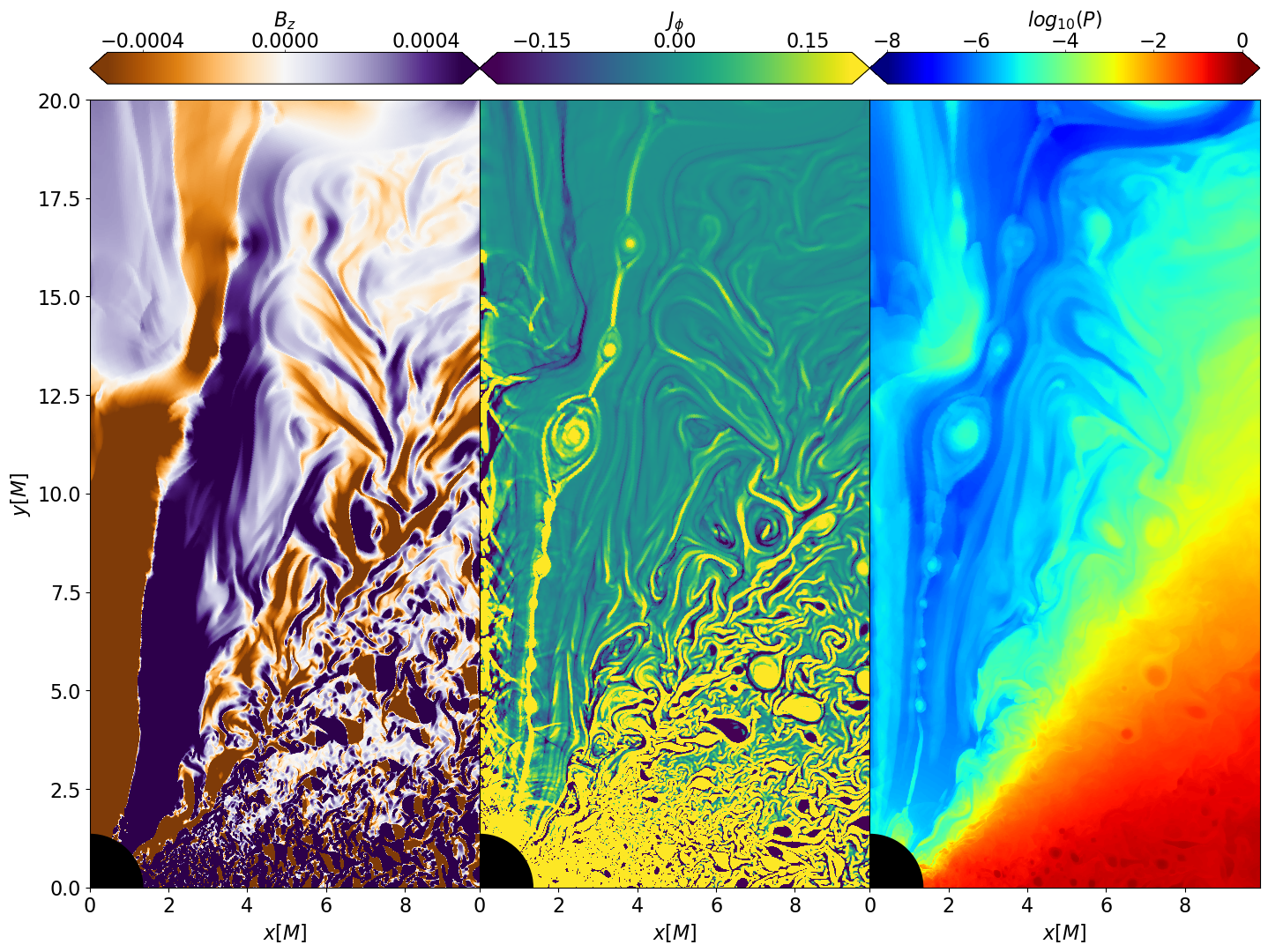}
        \includegraphics[width=1.3\columnwidth]{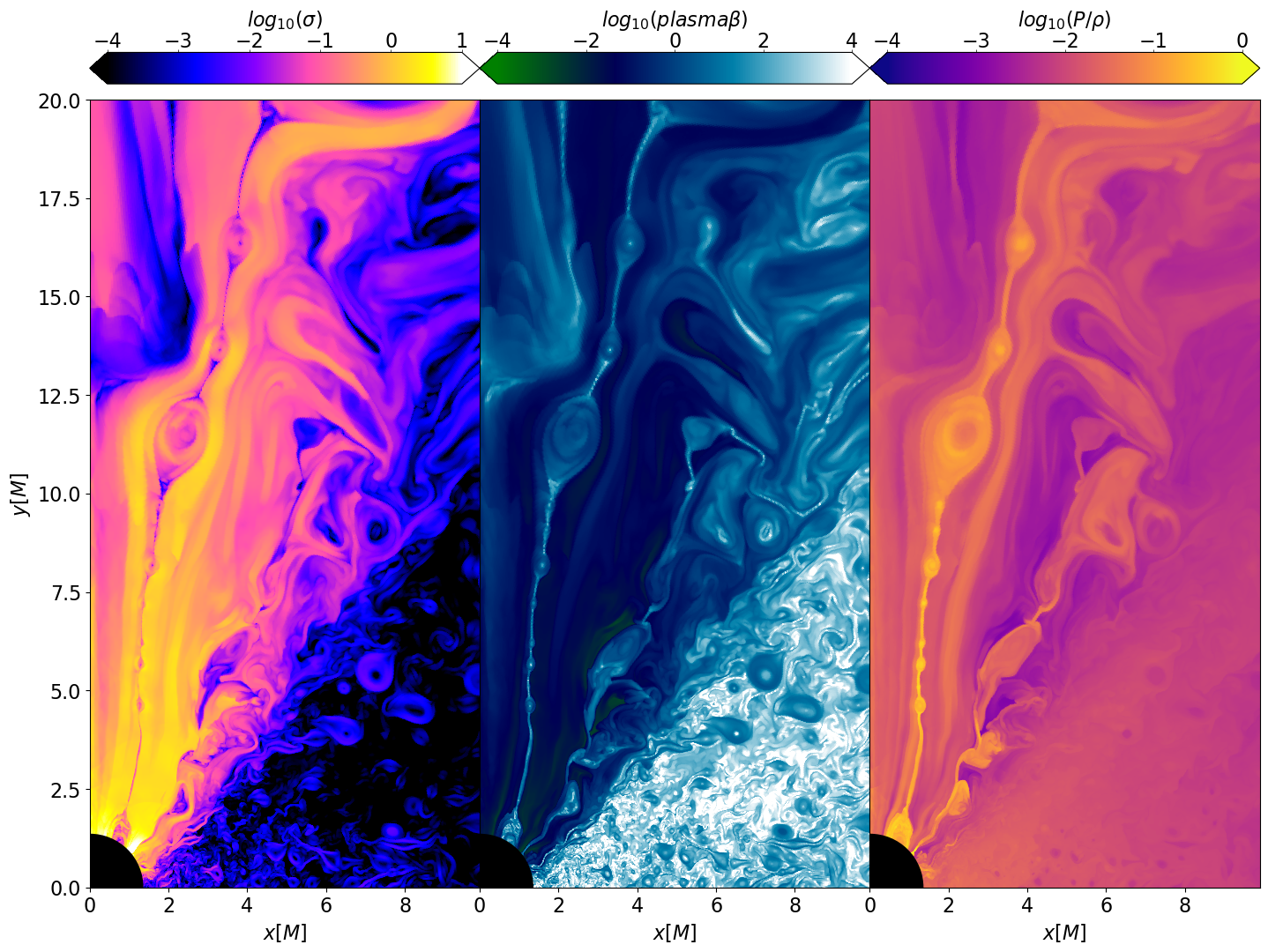}
        \caption{Upper panel (from left to right): magnetic field $B_z$ component, toroidal current and density.
        Lower panel (from left to right): magnetization, plasma~$\beta$ and dimensionless temperature
        $\Theta_p$.  By setting
        limits in these parameters, we define the red regions in
        Fig.~\ref{fig:env_cs}.}
        \label{fig:data404_polyplot}
    \end{figure*} 
    
\subsection{Radiation model}
	\label{sec:radiation}
In this subsection we discuss and present in detail the model that we used
	for the thermal and non-thermal radiation expected from plasma in
	the accretion flow. We employ an approximate "proxy" model for radiation which integrates the thermal emissivity across the entire disk.
The validity of this method has been tested in the past against  General Relativistic Radiative Transfer (hereafter GRRT) calculations 
	\citep{porth2019}. These comparisons demonstrated a close correspondence between the proxy method's results, the detailed GRRT calculations, and the observed variability in the lightcurve.   \citet{scepi2022sgr} employed a similar strategy to investigate flares originating from current sheets and plasmoids for MAD models.

	The total luminosity $L_{\nu}$ emitted by the fluid at a specific frequency
 $\nu$ (in units of ergs per second per Hz), is calculated  based on the following equation:
	\begin{equation}
    \begin{split}
			L_{\nu}=\int_{0}^{2\pi}\int_0^\pi\int_{r_h}^{r_{max}}
			e^{-\tau_{\nu}} \times j_\nu(B,p,\rho) \times g_{\rm redshift}^3 \\ \sqrt{-g}\ {\rm d}\phi {\rm d}\theta {\rm d}r
   \label{Lnu}
   \end{split}
	\end{equation}
	where \(j_\nu\) is the emissivity of thermal or non-thermal processes at
	a specific frequency\footnote{There seems to be some confusion in the literature in the definition of `emissivity'. Emissivity usually expresses the effectiveness of the surface of a material in emitting energy as thermal radiation. In the recent astrophysical literature, however, emissivity $j_\nu$ is defined as the energy loss per unit time per frequency per unit volume (e.g. \citealt{leung2011numerical,ghisellini2013,scepi2022sgr}). Note that \citealt{rybicki_lightman} define $j_\nu$ as the `emission coefficient', i.e. the energy loss per unit time per frequency per unit volume per steradian.} and $\tau_\nu$ is the opacity of the fluid to a given process at frequency $\nu$.
 The terms $\sqrt{-g}\ {\rm d}\phi {\rm d}\theta {\rm d}r$ represents the infinitesimal volume element $dV$ for each grid point.
 Close to the black hole radiation in specific is suppressed by the gravitational redshift 
    term  ($g_{\rm redshift}^3$; \citealt{viergutz1993}), where $g_{\rm redshift}=\sqrt{ \Sigma \Delta / A}$. Here, $A=(r^2+a^2)^2-a^2 \Delta \sin^2 \theta, \Sigma = r^2+a^2\cos^2 \theta$ and $\Delta = r^2-2rM+a^2$.
  The parameter $a$ corresponds to the spin of the BH, while $M$ represents its mass, which is fixed at $1$ in these forms (an estimation of $g_{\rm redshift}$ for $r=2r_g$, $\theta=\pi/4$ and $a=0.93$ is approximately on the order of $10^{-1}$).
  The observable radiation flux is given by
	\(F_\nu=L_\nu/4\pi D^2\), where the distance from the black hole is
	$D=8.277$ kpc (\citealt{abuter2022mass}). The integration of the luminosity works best when the viewing angle is nearly face on, as we are not taking into account the effects of Doppler beaming and gravitational lensing. 
The difference of such a method compared with a full GRRT calculation is of order unity
\citep{scepi2022sgr}.

At 230~GHz, the thermal electron synchrotron radiation comes predominantly from  the disk 
region, namely $\pi/3\lsim\theta\lsim 2\pi/3$.
In order to calculate the
synchrotron emission of a thermal distribution of plasma, we utilize the
formula developed by \citet{leung2011numerical} (see also \citealt{pandya2016}). At 230~GHz, the emissivity of a fluid element in the disk is given by the equation
	\begin{equation}\label{emissivity_thermal}
			j_{\nu}= n_e\frac{\sqrt{2}\pi e^2 \nu_s}{6\Theta_e^{-2}c}  X e^{-X^{1/3}}
	\end{equation}
$n_e$ is the electron number density and $X\equiv\nu/\nu_s$, where $\nu_s\equiv (2/9)\nu_L\Theta_e^2\sin{\lambda}$ ($\lambda$ is the particle pitch angle with respect to the direction of the magnetic field), and
$\nu_L\equiv eB/(2\pi m_ec)$ is the Larmor frequency.  We use $\Theta_e$ to denote 
the dimensionless temperature of the electrons defined as
	\begin{equation}\label{theta_e}
			\Theta_e= \frac{m_p}{m_e}\frac{T_e}{T_p}\frac{P}{\rho}
	\end{equation}
where $m_p, m_e$ are the masses of the proton and electron respectively, it is important to mention that the composition of the fluid is assumed to be electron-proton plasma. 
The ratio $T_p/T_e$ is calculated from the prescription of \cite{moscibrodzka2016general} which describes the 
  electron-to-proton coupling in low 
	(disk) and  high magnetized regions (funnel), namely
		\begin{equation}\label{ratio_temperature}
			\frac{T_p}{T_e}=R_{high}\frac{\beta^2}{1+\beta^2}+R_{low}\frac{1}{1+\beta^2}\ ,
	\end{equation}
		where $R_{high}$ and $R_{low}$ are free parameters that determine the heating
	ratio of electrons. In our calculations we set
	$R_{high}=20$ and $R_{low}=1$, since we do not focus on a large parametric
	investigation of flares from such configurations. 
 In \citet{dihingia2023b}, a comparison was made regarding the prescription of temperature ratios proposed by various studies (\citealt{moscibrodzka2016general}; \citealt{dihingia2023a}; \citealt{meringolo2023}). 
The results indicate that the model presented by \citet{moscibrodzka2016general} offers the most accurate approximation of temperature ratios within the disk region. 
The electron temperature in a GRMHD simulation can be determined through two temperature simulations that independently evolve this quantity \citep{Jiang2023}. Such studies highlight the importance of self-consistently evolving the electron temperature to accurately explore the 
thermal synchrotron radiation \citep{Jiang2024}.

The thermal radiation absorption term $e^{-\tau_{\nu}}$ is important only inside the disk.  
The optical depth is defined as $\tau_{\nu}=\int_{\pi/3}^{\theta_0}a_{\nu,th} ds$ where $ds=\sqrt{g_{\theta\theta}} \, d\theta$ ($\pi/3 \leq \theta_0 \leq 2\pi/3$ 
are the latitudinal limits of the disk). In the case of local thermal equilibrium, the source function of the emitting material is given by the Planck function $B_{\nu}$. 
Then, the absorption coefficient of the material can be expressed as
\begin{equation}\label{absorptivity_thermal}
  a_{\nu,th}=j_{\nu}/B_{\nu}\ ,
\end{equation}
where
\begin{equation}\label{planck_function}
    B_{\nu}=(2h\nu^3/c^2)[\exp(h\nu/kT_e)-1]^{-1}
\end{equation}
is the Planck function. 
So, the total luminosity for thermal radiation at each grid point $(r_0,\theta_0)$ is calculated as follows: First, we determine the emissivity at the point. 
Next, we compute the absorption term at all grid points $(r,\theta)$ where $r = r_0$ and $\pi/3~\leq~\theta~\leq~\theta_0$. 
The sum of absorption coefficients ($a_{\nu,th}$) at these points provides us with the value of the optical depth  $\tau_\nu$ at $(r_0,\theta_0)$.
In this way, it is like integrating along light rays perpendicular to the equatorial plane.
 
 The non-thermal emissivity at 2.2~$\mu$m, which is attributed to the synchrotron radiation of relativistic electrons, is given by (\citealt{rybicki_lightman},\citealt{ghisellini2013})\footnote{The approximation by \citealt{ghisellini2013} ssumes that the min electron
Lorentz factor is one and the maximum electron Lorentz factor is infi-
nite. This implies that there is no high energy cut off in the emissivities,
i.e. the spectral slope is the same at high energies. This is not 100%
correct, and as a consequence, we are overestimating the high energy
fluxes which includes the NIR, and certainly the X-rays.}
	\begin{equation}\label{emissivity_nonthermal}
			j_{\nu}= \frac{3\sigma_T c K U_B}{16\pi\sqrt{\pi}\nu_L}\left(\frac{\nu}{\nu_L}\right)^{-\frac{p-1}{2}}
	\end{equation}
	where $U_B=B^2/(8\pi)$ is the energy density of the magnetic field.
	One needs also to determine the non-thermal electron distribution
	$n(\gamma)=K\gamma^{-p}$, so $K$ is the electron number density multiplied by the efficiency $\epsilon$ of the radiation mechanism, and $p$ is the slope of the distribution. 

In the case of  non-thermal radiation the integration of Eq.~(\ref{Lnu}) is performed along the radial direction. In the funnel region (see e.g. colored region in Fig.~\ref{fig:env_cs}) , this is like integrating along rays perpendicular to the equatorial plane. The optical depth in this case is 
	$\tau_{\nu}=\int_{r}^{r_{max}}a_{\nu,nth} ds$ where $ds=\sqrt{g_{rr}}\,dr$.
	$a_{\nu,nth}$ is the dimensionless absorptivity which for the non-thermal process is defined as 
\begin{equation}\label{absorptivity_nonthermal}
    a_{\nu,nth}=\frac{\sqrt{\pi}e^2K}{8m_ec}\nu_L^\frac{p+2}{2}\nu^{-\frac{p+4}{2}}f_a(p)
\end{equation} 
 (\citealt{ghisellini2013}; \citealt{pandya2016}), where the function $f_a(p)$ is a product of $\Gamma$ functions, approximated by 
\begin{equation}
\label{gamma_product}
f_a(p)\thicksim 3^{\frac{p+1}{2}}\left(\frac{1.8}{p^{0.7}}+\frac{p^2}{40}\right),
\end{equation}
also in this case, the total luminosity for non-thermal radiation at each grid point is calculated as follows: 
First, we determine the emissivity at the point $(r_0,\theta_0)$. 
Next, we compute the absorption term at all grid points $(r,\theta)$ where $r_0 \leq r \leq r_{max}$ and $\theta=\theta_0$. 
The sum of absorption coefficients ($a_{\nu,nth}$) at these points provides us with the value of the optical depth  $\tau_\nu$ at $(r_0,\theta_0)$.

A GRMHD simulation does not have the
necessary micro-physics to follow the acceleration of particles and extract the electron distribution function or the efficiency 
for microscopic processes occurring within the plasma.
In order to incorporate
how energy is transferred from the magnetic field to the plasma, we used two
different post-processing formulas from results of PIC simulations. The first
approach is from a local investigation of idealized current sheets \citep{ball2018slope}. The second one is from an investigation of
microphysical properties of special-relativistic turbulence \citep{meringolo2023}. The first approach was applied to the current sheets in the funnel 
region above the disk boundary, and the second one in the vicinity of the disk boundary. The results of such local investigations give us the opportunity to approximate the properties of the
accelerated particles that will eventually radiate through the local plasma characteristics. Let us first define the non-thermal acceleration efficiency $\epsilon$ as the fraction of the kinetic energy carried by the non-thermal particles to the kinetic energy of the total electron population,
    \begin{equation}\label{eq:efficiency_definition}
       \epsilon=\frac{\int_{\gamma_{pc}}^{\infty}(\gamma-1)[\frac{dN}{d\gamma}-f_{MB}(\gamma,\Theta_e)]d\gamma}{\int_{\gamma_{pc}}^{\inf}(\gamma-1)\frac{dN}{d\gamma}d\gamma}\ ,
    \end{equation}
    where $\gamma_{pc}$ denotes the peak of the spectrum and $f_{MB}$ is a relativistic Maxwellian fitting function.

According to \citet{ball2018slope}, Eq.~(\ref{eq:slope_Ball})
gives us the slope of the spectrum with respect
to the magnetization $\sigma$ and plasma~$\beta$.
\begin{equation}\label{eq:slope_Ball}
		p= A_p + B_p\tanh{(C_p\ \beta)}
\end{equation}
    where $A_p=1.8+0.7/\sqrt{\sigma}$, $B_p=3.7\sigma^{-0.19}$, and $C_p=23.4\sigma^{0.26}$.
    Moreover,  the electron non-thermal efficiency with respect to the magnetization ($\sigma$) and plasma~$\beta$ is
    \begin{equation}\label{eq:efficiency_Ball}
	\epsilon=A_\epsilon + B_\epsilon \tanh(C_\epsilon\beta)\ ,
    \end{equation}
    where $A_\epsilon=1-1/(4.2\sigma^{0.55}+1)$, $B_\epsilon=0.64\sigma^{0.07}$, and $
    C_\epsilon=-68\sigma^{0.13}$.
    On the other hand Eq.~(\ref{eq:slope_meringolo}) approximates 
    the slope $p$ (for Eq.~(\ref{eq:slope_meringolo}) it holds that $p=k-1$) with respect to the same parameters based on 
    \citet{meringolo2023}.  
    \begin{equation}\label{eq:slope_meringolo}
	    k = k_0 + \frac{k_1}{\sqrt{\sigma}} + k_2\sigma^{-6/10}\tanh{(k_3\beta\sigma^{1/3})}\ ,
    \end{equation}
    where $k_0=2.8$, $k_1=0.2$, $k_2=1.6$, and $k_3=2.25$.
    The efficiency of particle acceleration in turbulent plasmas is equal to
    \begin{equation}\label{eq:efficiency_meringolo}
        \epsilon = e_0 + \frac{e_1}{\sqrt{\sigma}}+e_2\sigma^{0.1}\tanh[e_3\beta\sigma^{0.1}]\ ,
    \end{equation}
    where $e_0=1$, $e_1=-0.23$, $e_2=0.5$ and $e_3=-10.18$.
    
    In the upcoming section \ref{sec:determination_cs}, we will elaborate on how we determine current sheets and their associated parameters, such as magnetization and plasma~$\beta$ in their 
    vicinity.  The reason we are interested in the latter is because dissipative current sheets are fed with plasma from their environment. These conditions determine the strength of the non-
    ideal electric fields that accelerate particles and the rate of energy dissipation. All empirical relations that describe properties of the accelerated particle populations refer to the $
    \sigma$ and plasma $\beta$ parameters in the upstream regions of the current sheets. 
    Once these areas of interest are identified, these formulas can then be applied to determine the slope of the distribution of accelerated electrons within the current sheets or turbulent regions at the disk boundary.

\subsection{Determination of current sheets and their environment}
    \label{sec:determination_cs}

    PIC simulations that investigate current sheets start with a specific magnetization, a specific
    plasma~$\beta$ and an initial current layer in the middle. As the
    magnetic field evolves magnetic reconnection events take place, magnetic energy is dissipated, and
    particles are accelerated from the thermal pool to higher energies. Eqs.~(\ref{eq:slope_Ball}) and (\ref{eq:slope_meringolo}) are fitting models of the results of
    such PIC simulations which can give us the slope of the spectrum of the
    non-thermal electrons as functions of GRMHD quantities. It is important to emphasize that the parameter values that must be used in
    these formulas are roughly those set as initial
    conditions in a PIC simulation. As it turns out, these values are
    maintained during the calculation only in the environment surrounding the
    current sheets.

\begin{figure}
       \centering
       \includegraphics[width=0.8\columnwidth]{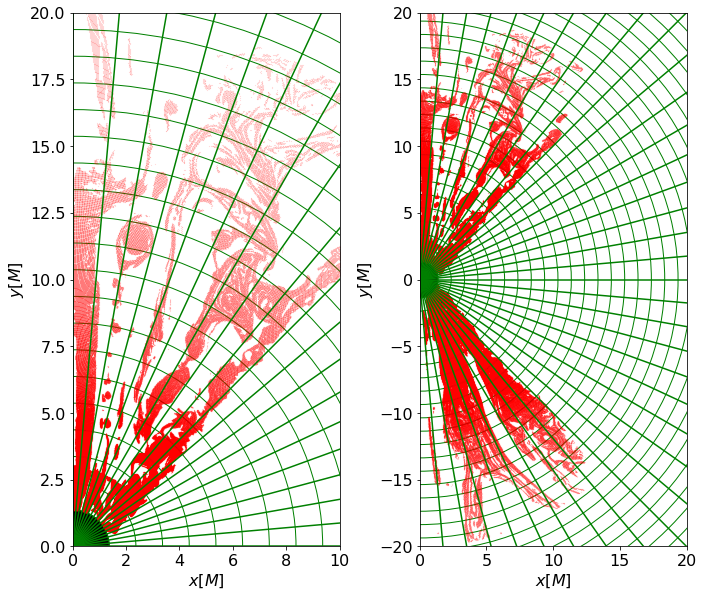}
       \caption{ Determination of plasma (red color) in the reconnection region and places of high magnetic turbulence.  They are selected after applying limits 
       on toroidal current, magnetization, plasma~$\beta$, density and dimensionless temperature. The green grid corresponds to the discretization for the purpose of parameter averaging. Same 
       snapshot as  in Fig~\ref{fig:data404_polyplot}. }   
       \label{fig:env_cs}
    \end{figure}

    So, in order to use eqs.~(\ref{eq:slope_Ball}) and (\ref{eq:slope_meringolo}) it is necessary to constrain the current sheet and its surrounding environment.
    First, it is useful to distinguish the regions where current sheets form from the polarity reversal of the magnetic fields (Fig.~\ref{fig:data404_polyplot}, reversal of $B_z$). 
    We utilize five variables to determine the current sheets, which are as follows: toroidal current $J_{\phi}\defeq(\nabla \times B)_{\phi}$, density $\rho$, magnetization $\sigma$, plasma~$\beta$ and 
    dimensionless temperature $\Theta_p$ (normalized to $mc^2$ which is equal with $P/\rho$).
    The remaining plots in Fig.~\ref{fig:data404_polyplot} provide limits for each parameter that defines the exact region where a current sheet occurs.
    As we see in Fig.~\ref{fig:data404_polyplot}, a current sheet is characterized by high values of toroidal current, high values of density, low values of magnetization, high values of plasma~$\beta$ and high values of dimensionless temperature.
    Thus, we set the following limits for each parameter: $|J_{\phi}|>10^{-4}$, a density
    floor (cutoff) at $\rho_{cut}=2\times10^{-5}$, a magnetization ceiling cutoff at
    $\sigma_{cut}=10$, a plasma~$\beta$ floor at $\beta_{cut}=10^{-3}$ and finally
    a temperature floor at $\Theta_p=10^{-3}$. By combining these constraints, we obtain the red regions in Fig.~\ref{fig:env_cs} which are clearly associated with the current sheets we identified.
    Additionally, the plasma in these red-colored regions has undergone the reconnection process. Efficient particle acceleration occurs in regions with high magnetization and 
    low plasma~$\beta$ \citealt{ball2018slope}. Such regions develop only in the funnel region, so we applied the above methodology exclusively to the plasma within this region.

 Following this procedure we have identified the current sheets developed in the simulation. The next step involves characterizing the current sheet environment in terms of averaged quantities  in order to use Eq.~(\ref{eq:slope_Ball}) and (\ref{eq:slope_meringolo}) for 
 the plasma within the current sheet. To achieve this, a discretization method is employed, wherein the region is divided into cells with a radial length of 1 $r_g$ and a meridional extent of 10 degrees in the $\theta$ direction.  In each cell that contains
    part of a current sheet (red region) we calculate the mean values of
    magnetization $\sigma$ and plasma~$\beta$ of their environment (white region) over 2 cells to the right and 2 to the left in the
    $\theta$ direction, namely
    \begin{equation}\label{mean_magn}
       \overline{\sigma}=\frac{1}{4}\sum_{cell=-2}^{cell=+2}\sigma 
    \end{equation}
    \begin{equation}\label{mean_plasmab}
       \overline{\beta}=\frac{1}{4}\sum_{cell=-2}^{cell=+2}\beta
    \end{equation}
    Subsequently, the slope of the non-thermal electrons inside the current sheet or in the turbulent plasma in each cell is computed using equations Eq.~(\ref{eq:slope_Ball}) and (\ref{eq:slope_meringolo}) with parameters values as
    $\sigma~=~\overline{\sigma}$ and $\beta~=~\overline{\beta}$. The plasma assumed to be accelerated corresponds to the region depicted in red in Fig. \ref{fig:env_cs}.


\section{Results}
    \label{sec:results}
    In the next two subsections, we present our results on the thermal and non-thermal 
    radiation. The first part covers calculations at 230 GHz (thermal component), which 
    help calibrate the physical parameters for Sgr~A*. Further calculations provide 
    information about the model's variability compared to observations. Subsequently, we 
    present results from the non-thermal radiation calculations at 2.2~$\mu$m, including 
    the resulting light curve, an analysis of flares, and a comparison with the GRMHD 
    simulation results.

\subsection{Radiation at 230~GHz} 
\begin{figure}
       \centering
       \includegraphics[width=0.7\linewidth]{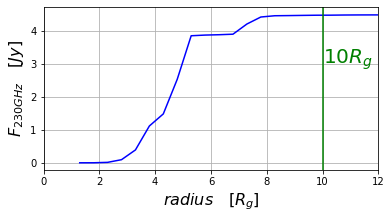}
       \caption{ Cumulative 
       230~GHz flux from all angles $\theta$ for radii less than or equal to $r$. }
       \label{fig:Radial_angle_profile_Flux}
    \end{figure}

    Observations of Sgr~A* at 230~GHz have been conducted by various
    telescopes, including the Submillimeter Array (SMA), the Atacama Large
    Millimeter/submillimeter Array (ALMA), and notably, the Event Horizon
    Telescope (EHT). The latter, in particular, has provided a wealth of
    high-resolution images from the innermost region of the accretion disk
    onto Sgr~A* itself, presenting an invaluable resource for the scientific
    community (\citealt{akiyama2022I}a-f).
    Thus there exists a plethora of data on the properties of the lightcurve
    at that particular frequency. Therefore, utilizing much of this data, our main goals 
    are to reproduce the main features of the light curve at 230~GHz (amplitude and variability) by calibrating the various physical parameters from dimensionless code units to actual physical units.
When we combine how the flux changes with distance and angle (as
illustrated in Fig.~\ref{fig:Radial_angle_profile_Flux}), it becomes clear
that the majority of the radiation is emanating from a distance of
approximately $10~r_g$.  This underscores that the events
we are studying occur in extremely close proximity to the black hole, and
consequently, the critical data points we are interested in are highly
concentrated in this area.

\begin{figure}
   \centering
   \includegraphics[width=\columnwidth]{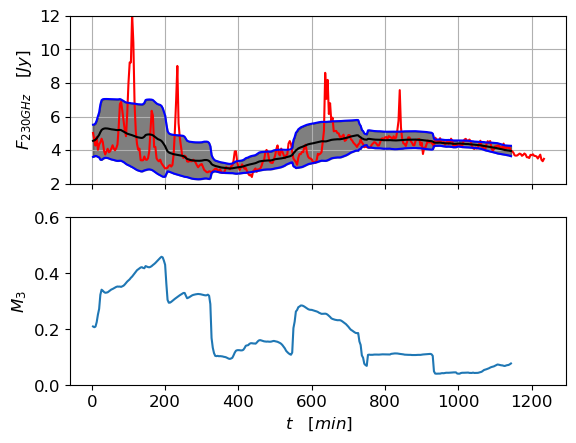}
   \caption{Top panel:  the lightcurve at 230~GHz (red line), the moving average of the lightcurve ($m_3$, in a time window of 3 hours, black line) and the 1$s_3$ standard deviation (grey area).  Lower panel: The measure of variability the lightcurve $M_3$ defined as
$M_3=s_3/m_3$.}
   \label{fig:Flux_230GHZ}
\end{figure}
\begin{figure}
   \centering
   \includegraphics[width=0.9\columnwidth]{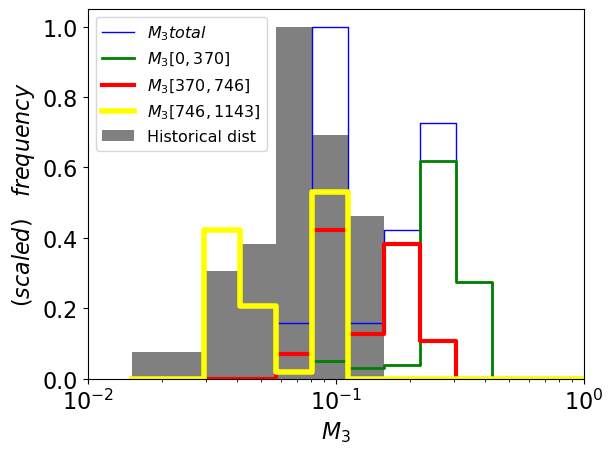}
   \caption{Distribution of $M_3$ of the lightcurve at 230~GHZ (blue)
compared with historical data from observations of Sgr~A* (grey).
We have also cut the lightcurve and respectively $M_3$ in
three time intervals, first for $t\in[0,370]$~min (green), second
for $t\in[370,746]$~min (red) and third for$t\in[746,1143]$~min
(yellow).}
   \label{fig:hist_M3}
\end{figure}

    As we know from observations (\citealt{wielgus2022millimeter}) the
    radiation flux of Sgr~A* at 230~GHz  is approximately stable. In order to
    turn the  parameters of the GRMHD simulation from code units to physical
    cgs units a conversion and calibration of variables is necessary. We
    define a scaling parameter S, that appropriately multiplies physical
    quantities such as magnetic field (B), pressure (P) and density ($\rho$)
    to convert values of physical quantities from geometric (code) units to
    cgs units. Obviously, the conversion leaves unchanged dimensionless
    quantities such as $\Theta_e$, $\sigma$, and $\beta$ 
    \begin{equation}\label{conversion_cgs_B}
        B_{cgs}=\sqrt{4\pi}cS^{1/2}B_{geometric}
    \end{equation}
    \begin{equation}\label{conversion_cgs_rho}
        \rho_{cgs}=S\rho_{geometric}
    \end{equation}
    \begin{equation}\label{conversion_cgs_P}
     	P_{cgs}=c^2SP_{geometric}
    \end{equation}
    Through an iterative procedure applied in various snapshots of the
    GRMHD calculation, the model limited the radiation flux in a range around
    4~Jy and finally produced the value of the scaling parameter $S$. After
    calculating the parameter $S$ we can calculate the radiation flux at all
    snapshots and finally produce the lightcurve at 230~GHz. The upper 
    panel of Fig.~\ref{fig:Flux_230GHZ} presents 
    the lightcurve at 230~GHz (red line).  Noticeably,  the radiation flux remains almost constant throughout the calculation period. The higher flux values observed in the beginning of the light curve gradually decrease towards the end of the evolution. 

    To investigate the variability of  the lightcurve
    we define  the moving  average radiation flux over a 3-hour
    period ($m_3$, black line in Fig.~\ref{fig:Flux_230GHZ}) and the standard deviation over the same duration ($s_3$, grey area in Fig.~\ref{fig:Flux_230GHZ}). The lower
    panel of  Fig.~\ref{fig:Flux_230GHZ} shows the spread of $M_3$ where $M_3=s_3/m_3$. $M_3$
    is a measure of the variability of the lightcurve and can be compared
    with historical observations from Sgr~A* (Fig.~\ref{fig:hist_M3}).

Sgr~A* lightcurve shifts and changes over time due to a combination of
stellar winds and turbulence on various scales. Longer changes come from
variations in the stellar winds, especially near the S stars. Shorter changes
happen because of turbulence closer to the center.  
The spread of $M_3$ in Fig.~\ref{fig:Flux_230GHZ} exhibits a downward trend, suggesting that if we keep running the same simulation for further time, the flare variability will eventually stabilize at lower levels. Fig.~\ref{fig:hist_M3}
is a histogram of $M_3$  that provides a broader view derived from observing Sgr~A*, the
historical data is represented in gray (see
\citealt{wielgus2022millimeter}). The other distributions are 
derived from the simulation results for different time windows.
It is important to keep in mind that the yellow line distribution  
represents the last part of the $M_3$ curve inside the time interval $t\in[746,1143]$~min. This
closely aligns with the observations and shows that a new
calculation of the same model for longer times will probably yield the
desired variability in the lightcurve.

\begin{figure}
   \centering
   \includegraphics[width=\columnwidth]{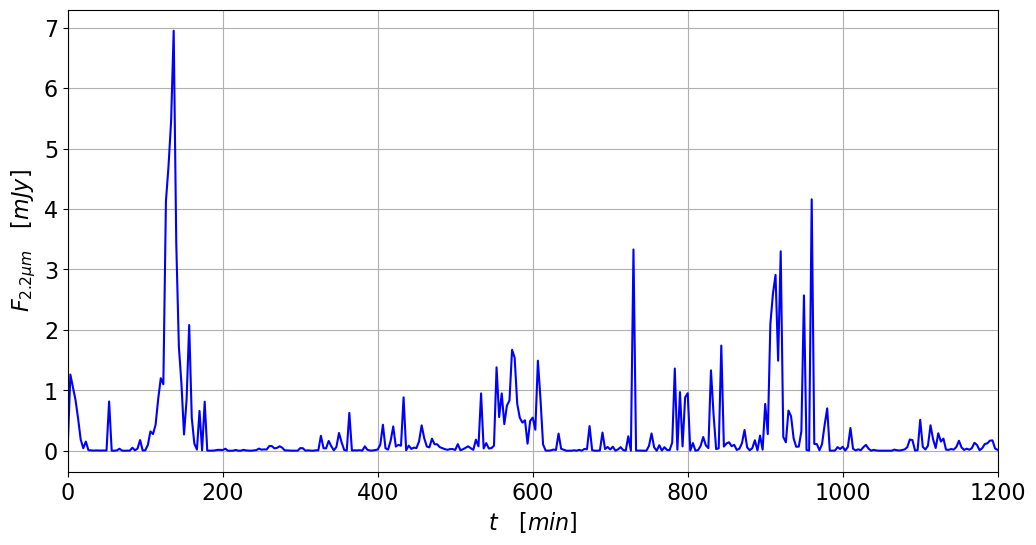}
   \caption{Lightcurve at $2.2\mu$m. Our model can produce one very bright flare around 7 mJy and another three that pass the 
   threshold of  1.5 mJy. }
   \label{fig:Flux_2.2}
\end{figure}
\begin{figure}
   \centering
   \includegraphics[width=1.\columnwidth]{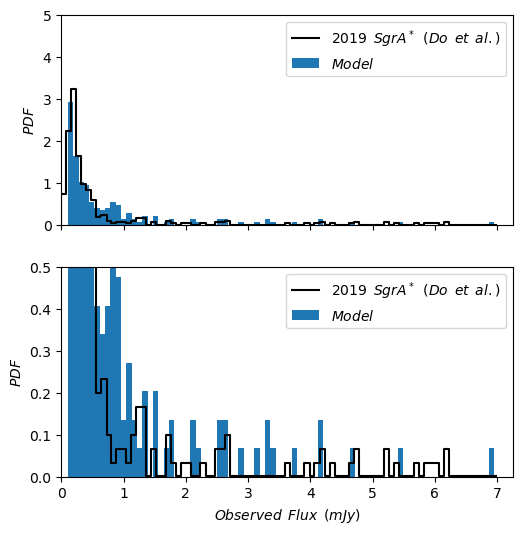}
   \caption{Histogram comparison between our model and observations from \citet{do2019unprecedented}.}
   \label{fig:hist_do}
\end{figure}

\subsection{Results at 2.2~$\mu$m (NIR flares)}
\subsubsection{Lightcurve and statistics}

In Fig.~\ref{fig:env_cs} we carefully marked the specific areas where
non-thermal radiation processes occur. These dynamic areas reside within
the funnel region and in the limit of the disk. There is an environment characterized by high magnetization and
low plasma~$\beta$ values which are important to activate the radiation mechanisms with 
sufficient efficiency. In Section~\ref{sec:determination_cs} we discussed in detail that it is this  local environment that characterises the process of acceleration for the plasma residing within current sheets.  

Using data from these regions, we calculate the light curve at $2.2~\mu$m, and our results are depicted in Fig.~\ref{fig:Flux_2.2}.
Notably, during quiescent states, the average flux is remarkably low, consistent with observations. Within the simulation, several flares occur, each typically exceeding a flux of 
$>1.5$ mJy \citep{Witzel2021, vonFellenberg2023}. Noteworthy, there are four flares during the simulation time that surpass this threshold, with an additional three flares marginally 
meeting the criteria for acceptance as flares.
One particularly significant radiation flare, with a flux of approximately 7 mJy and a longer duration than typical flares, stands out as a potential candidate within the observed 
limits for both flux and duration, resembling to the unprecedented flare reported in \citealt{do2019unprecedented}.

 In a theoretical comparison with observations we do not have a
flux limit that allows us to define what is considered as flare and what
not but it is obvious that the values of the flux in
Fig.~\ref{fig:Flux_2.2} rise up several orders of magnitude between a
quiescent and a flare state ($F \gsim 1.5$~mJy). Sgr~A* produces flares with
lower flux but also some very bright ones ($F>6$~mJy; 
\citealt{do2019unprecedented}) like those produced in our model. In Fig.~\ref{fig:hist_do} we see 
a comparison of the historical flux distribution in histogram for Sgr~A*
from \citealt{Witzel2018} as presented in \citet{do2019unprecedented} compared to the high flux flare observed in 2019. The probability 
of observing such a flare (>6mJy) was computed to be less than $0.05\%$
\citep{Witzel2018}.
Our model is capable of reproducing high-flux flares that cover the full range of observations.

According to Fig.~\ref{fig:hist_do} our model appears to have fairly good statistics in terms of the 
population of flares and in terms of their intensity. This is not the case for Sgr~A* flares generated from MAD models which seem to show a very large population of flares and and tend to overproduce the quiescent state flux (\citealt{dexter2020sgr,scepi2022sgr, White2022}). This is probably 
due to the very strong magnetic field they possess, and to the fact that the main source of 
the flares is the current sheet in  the equator which feeds high values of absorptivity in the density due to the $\Dot{M}$ (as seen characteristically in the paper of \citealt{scepi2022sgr}).

\begin{figure}
    \centering
    \includegraphics[width=0.9\columnwidth]{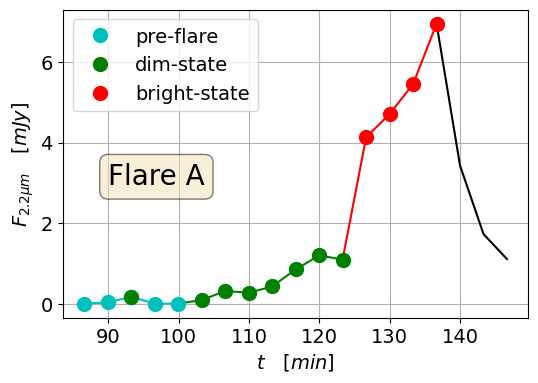}
    \centering
    \includegraphics[width=0.9\columnwidth]{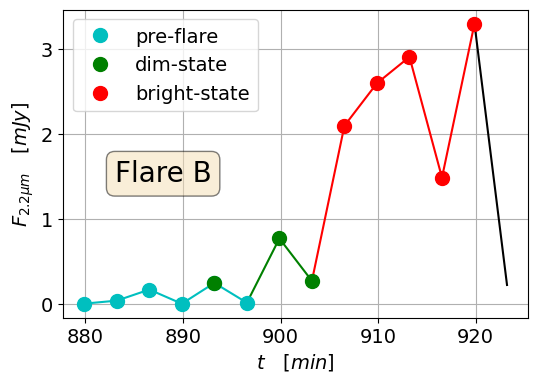}
    \caption{Part of the lightcurve at $2.2~\mu$m. In the upper plot, Flare A, for $t\in[86,150]$~min 
(the first bright flare with approximately 7 mJy radiation flux) and in the bottom plot, Flare B, for
$t\in[879,926]$~min (a bright flare with two peaks).
The duration of the flares is about 40-60 minutes, duration relevant to observations.
We can divide the lightcurve in three states: pre-flare, dim-state
and bright-state. This separation is based on the flux and spectral index}
    \label{fig:flare750_lc}
\end{figure}

\begin{figure}
    \centering
    \includegraphics[width=0.9\columnwidth]{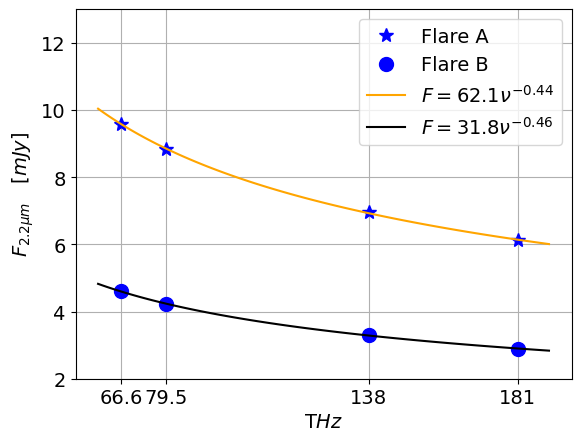}
    \caption{Radiation flux around $2.2~\mu$m. Specific to the frequencies 66.6 (M-
    band), 79.5 (L-band),
    138 (K-band) and 181 (H-band)~THz for Flare A (stars) and Flare B (circles). The
    fitting (black and orange) lines correspond to power-laws where the
    exponential gives us the spectral index of the spectrum, in our
    case $a=-0.44$ for Flare A and $a=-0.46$ for Flare B.}
    \label{fig:spectral_index_flare}
\end{figure}

\begin{figure}
    \centering
    \includegraphics[width=0.9\columnwidth]{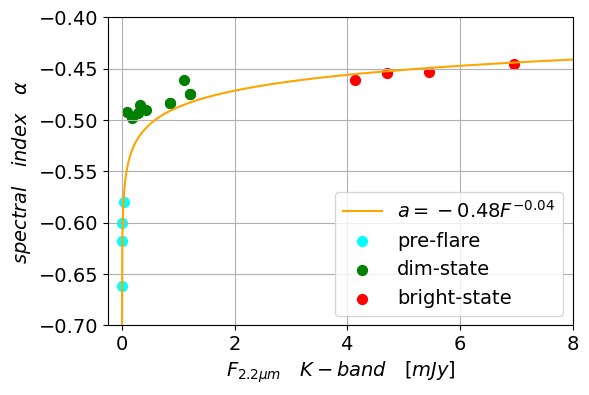}
    \centering
    \includegraphics[width=0.9\columnwidth]{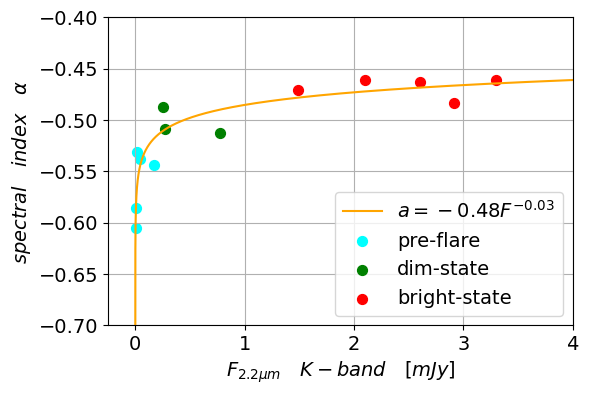}
    \caption{Spectral index vs Flux for the two flares (Flare A upper plot and Flare B bottom plot).
    The dots correspond to the snapshots that are marked also in Fig.~\ref{fig:flare750_lc}: 
    cyan color corresponds to pre-flare states, green color represents the dim-state, red 
    color represents the bright state of each flare. The orange lines are a fitting power-law 
    curve which can be compared with observations}
    \label{fig:spectral_index_vs_Flux}
\end{figure}

 Another crucial variable to consider is the cooling time of particles responsible for producing non-thermal radiation. Our method for computing the light curve involves each simulation snapshot radiating under the assumption of the fast cooling regime. Additionally, no cooling effects are considered, treating all radiation as if emitted instantaneously. The accuracy of this approach depends on the estimation of the cooling parameter.
    The cooling time of a particle with Lorentz factor $\gamma$ due to synchrotron radiation is given by the equation:
    \begin{equation}\label{eq:cooling_time}
       t_{syn}=\frac{\gamma m_e c^2}{4/3\sigma_TcU_B\gamma^2\beta^2}
    \end{equation}
    where the Lorentz factor $\gamma$ is  $\gamma=(\nu/\nu_L)^{1/2}$. Using typical values for  
    magnetic field B from our model, approximately 60~G for the region of interest calculating the cooling time from relation (\ref{eq:cooling_time}) 
    we obtain $ t_{syn}\sim 3$~min. This suggests that particles have enough time to radiate their  energy
    in a snapshot of the GRMHD simulation, given that simulation outputs occur every 10 M, which translates to approximately 200 seconds for Sgr~A*. Therefore, cooling effects do not significantly impact the calculation.

\begin{figure*}
   \centering
   \includegraphics[width=0.76\linewidth]{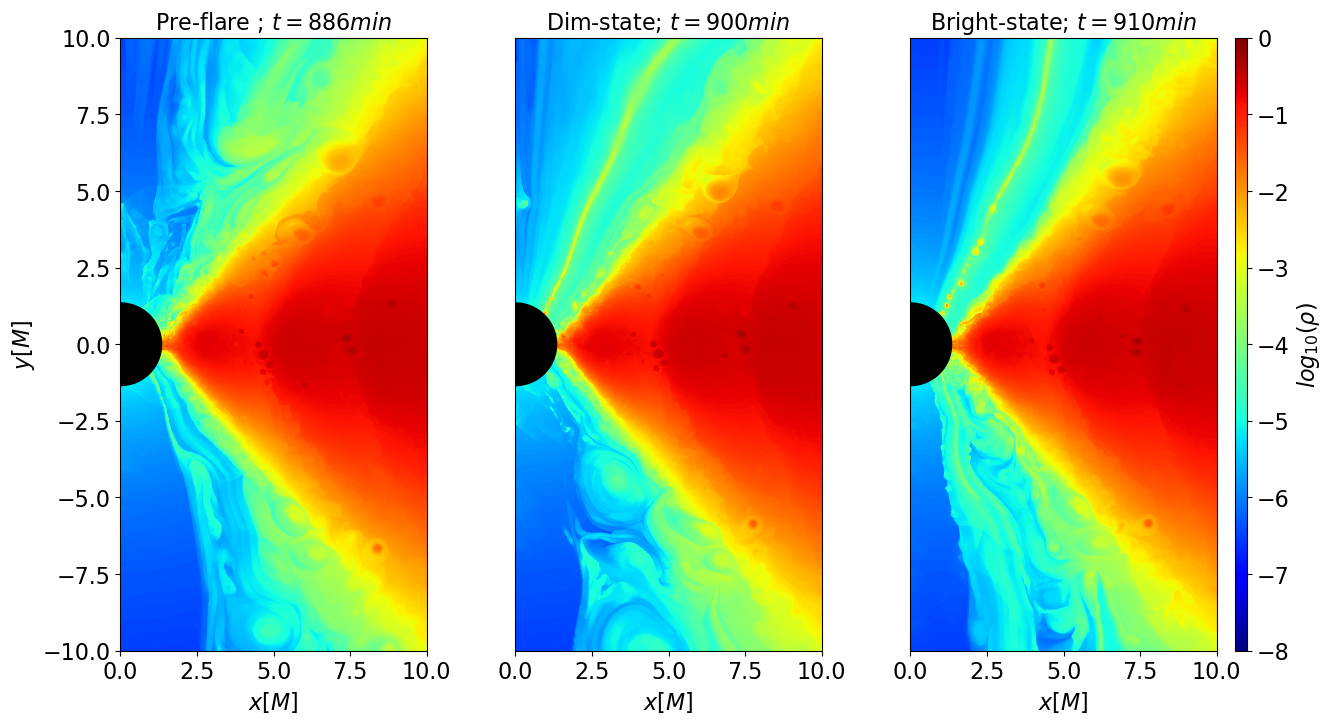}
   \centering
   \includegraphics[width=0.76\linewidth]{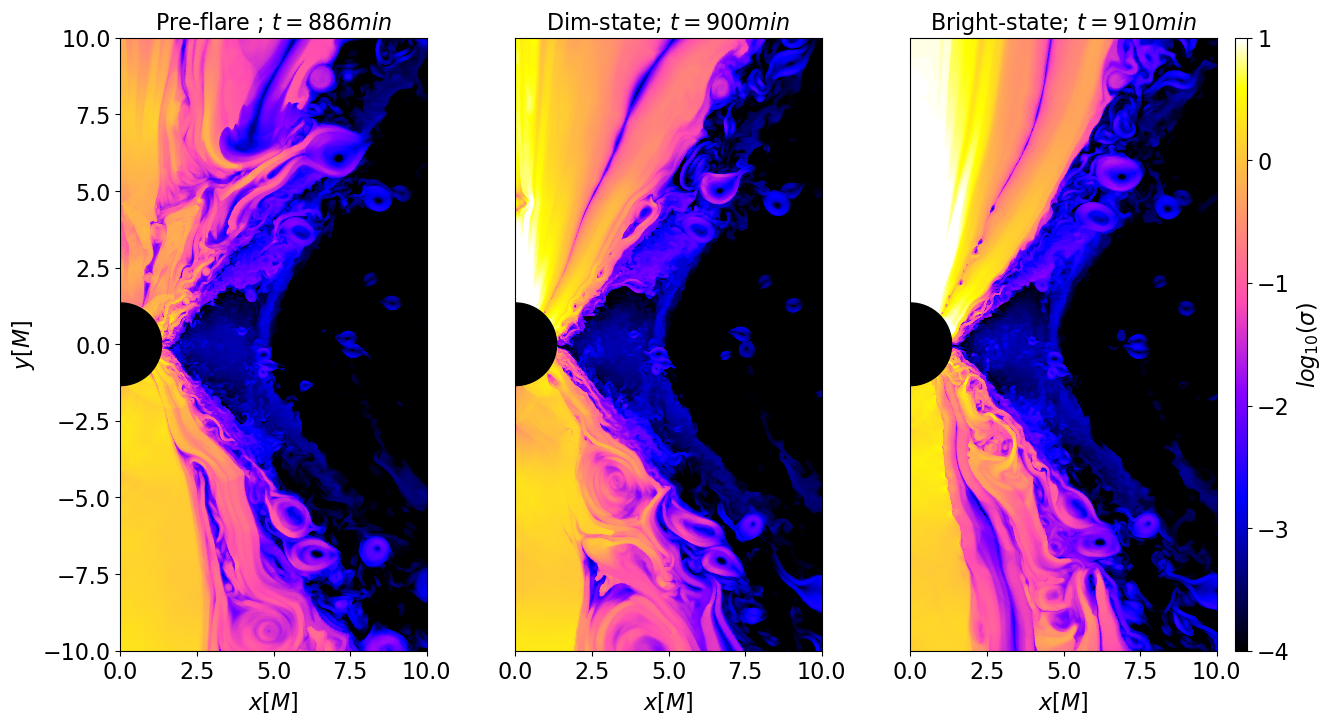}
   \centering
   \includegraphics[width=0.76\linewidth]{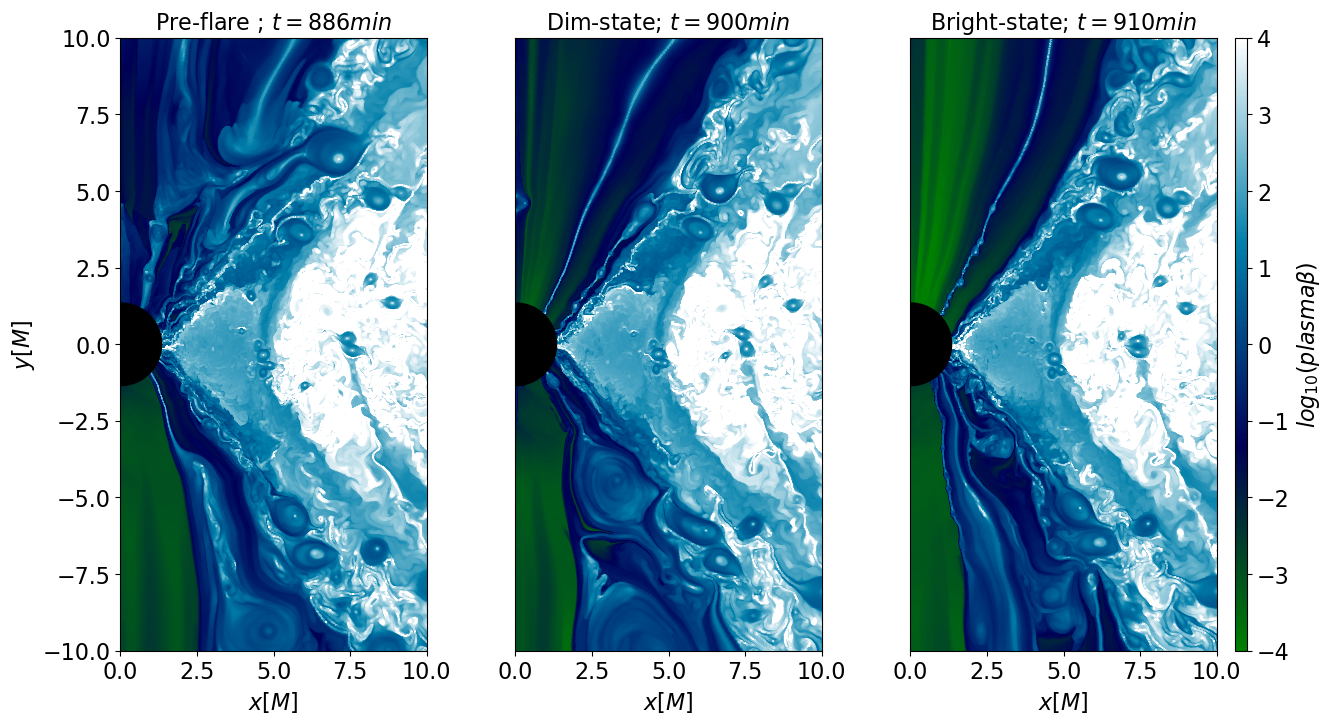}
   \caption{Top row: density of the second bright flare
   ($t\in[886,963]$~min) for its three states: pre-flare, dim-state,
   birght-state at snapshots $t=886$~min, $t=900$~min and
   $t=910$~min respectively. Middle row: magnetization ($\sigma$) at the same
snapshots. Bottom row: plasma~$\beta$ at the same snapshots. All panels refer to Flare B.}
   \label{fig:flare750_states}
\end{figure*}
\begin{figure}
     \centering
     \includegraphics[width=\columnwidth]{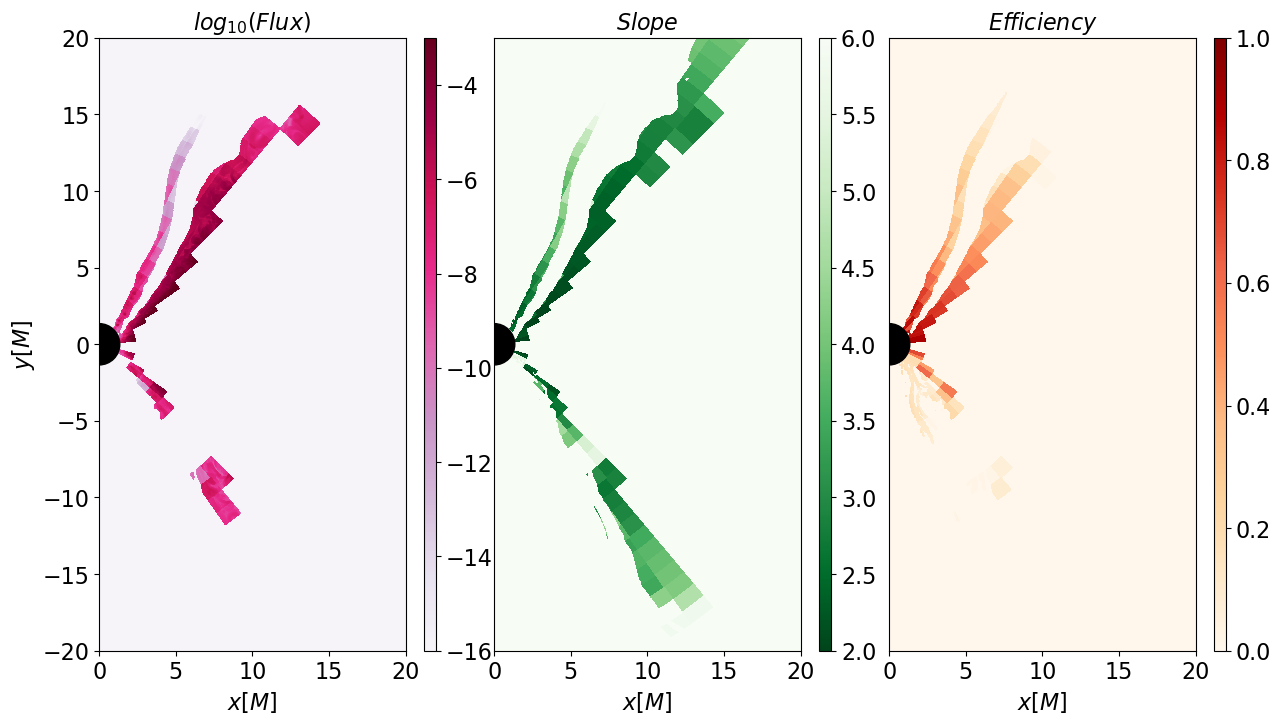}
     \caption{Radiation flux at 2.2 microns (in mJy),  slope $p$ and  efficiency $\epsilon$  in each grid point of the 
     	GRMHD simulation are shown in the left, middle and right panels, respectively. The plots illustrate the bright state of Flare B.}
     \label{fig:flux_slope_efficiency_data750}
\end{figure}

\begin{figure}
    \centering
    \includegraphics[width=\columnwidth]{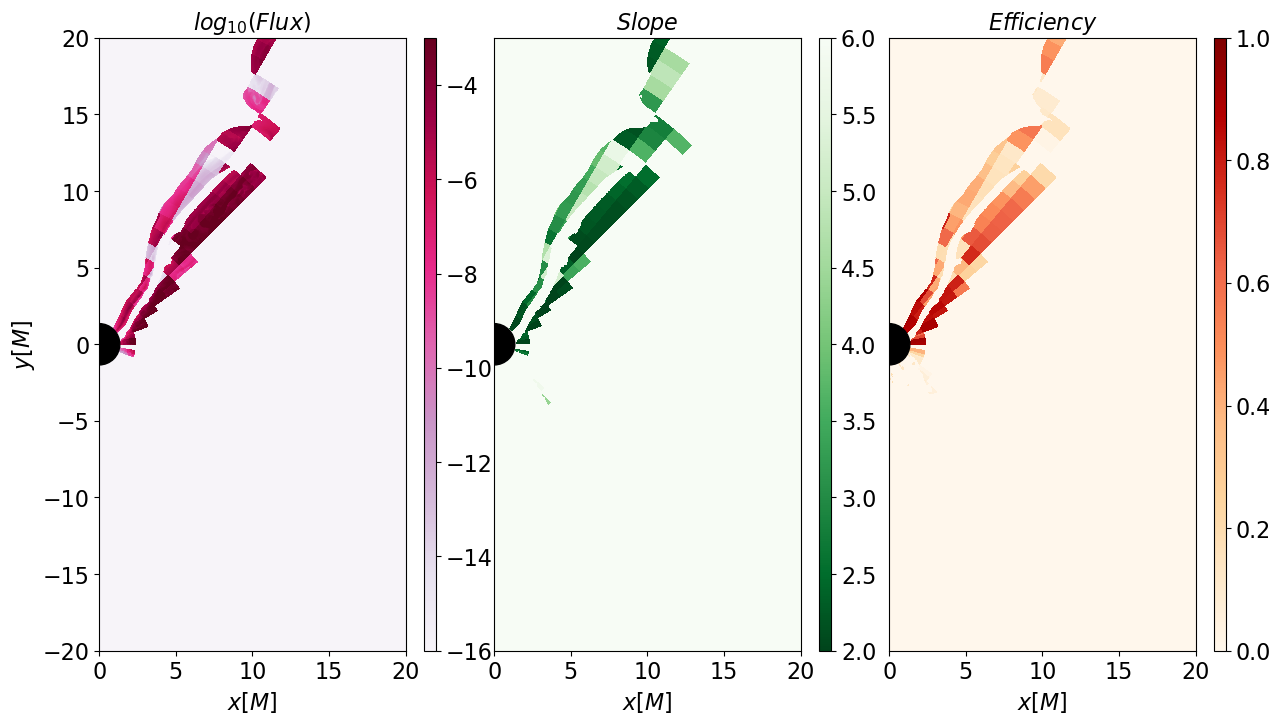}
    \caption{Same plots as Fig.~\ref{fig:flux_slope_efficiency_data750} but for  the bright 
     	state of Flare A.}
    \label{fig:flux_slope_efficiency_data443}
\end{figure}

\subsubsection{Properties of flares}
In the upper panel of Fig.~\ref{fig:flare750_lc} we zoom in on the first major flare within the time range $t\in [86,150]$, 
hereafter referred to as Flare A. Whereas, on the lower panel, we zoom on a second, smaller flare within the time range $t\in [879,926]$, 
hereafter referred to as Flare B. We examine how the properties of each flare change across three 
different stages: the pre-flare state, which occurs when the flux is several orders of magnitude 
smaller than 1~mJy; the dim-state, which occurs when the flux begins to increase 
(as shown in Fig.~\ref{fig:spectral_index_vs_Flux}, where the 
spectral index also increases) and finally, the bright-state, which occurs when the flux exceeds 1 mJy.
In both flares, we can see the specific points and their corresponding colors that represent the
three different states. Each point is derived from a specific snapshot of the GRMHD simulation. 
The first flare is characterized by a distinct peak, while the second flare has two peaks. 
Similar phenomena have been observed (\citealt{gravity2018flares}). 
The classification of points into the different flare states was based on flux 
and spectral index. The rationale for this classification is 
clearly illustrated in Fig.~\ref{fig:spectral_index_vs_Flux}.

Following the same calculation methods used for the non-thermal radiation, 
we estimate the radiation flux at three additional frequencies: 66.6, 79.5, 
and 181~THz, which correspond to the M, L, and H bands, respectively. The 
primary frequency of our calculation, $2.2~\mu$m (138~THz), corresponds to the K-band for Sgr A*.
For this calculation we used two snapshots at times $t=136$~min and $t=920$~min, one for each flare at 
the peak of the bright state. We thus extract the spectrum of our model for this flare state
(Fig.~\ref{fig:spectral_index_flare}).
The fitting curves in Fig.~\ref{fig:spectral_index_flare} can give us the spectral index 
(which defined as the exponent of
a power law like $F\propto \nu^{a}$) for each 
flare. The two values that we obtained ($a=-0.46$ and $a=-0.44$) are within the observed 
range (\citealt{ghez2005stellarb, ghez2005first, bremer2011near}). 


During a flare, both the flux and the spectral index simultaneously increase. We  categorize the points 
into three groups, distinguished by the colors used to represent them.
Fig.~\ref{fig:spectral_index_vs_Flux} shows the
evolution of the spectral index as the flares transition from the pre-flare
state to the dim-state and finally to the bright-state. The fitting power law curve (orange line)
approximates how the spectral index evolves as the flare transitions from one state to the next, which is comparable to observations (\citealt{bremer2011near}).  
The same 
colors were used to distinguish the points in Fig.~\ref{fig:flare750_lc}.

\begin{figure}
    \centering
    \includegraphics[width=\columnwidth]{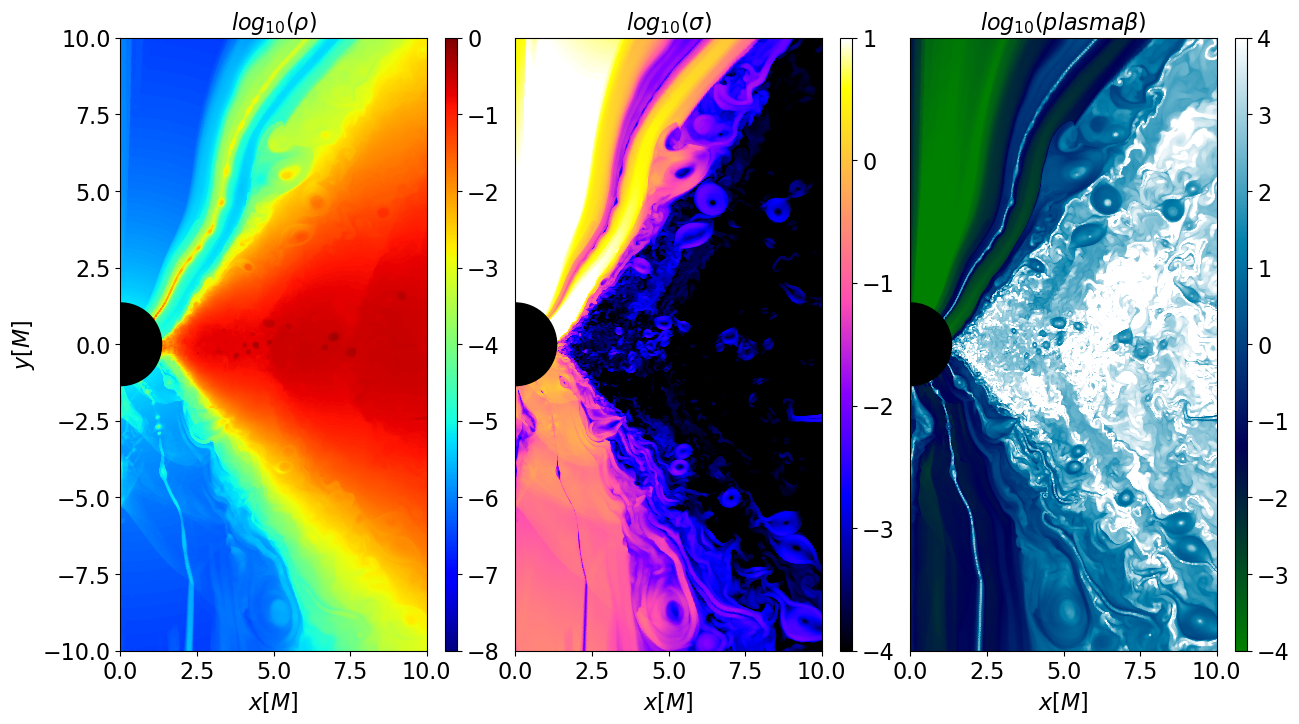}
    \caption{Density (left panel), magnetization $\sigma$ (middle panel) and plasma $\beta$ 
     	(right panel) for Flare A in the bright state.}
    \label{fig:rho_magn_plasmab_data443}
\end{figure}

\subsubsection{Flare states compared with GRMHD simulation}

As a final property of the flares we wanted to see how these peaks in the lightcurve relate to the 
GRMHD calculation from which the phenomena originated.
Therefore, we plotted the parameters $\rho$, $\sigma$ and plasma~$\beta$, 
which play crucial roles in determining the current sheet
(Fig.~\ref{fig:flare750_states}). The
plots refer to the three states of the second flare we investigated, specifically 
the snapshots at times  $t=886$~min (pre-flare), $t=900$~min
(dim-state), $t=910$~min (bright-state). The corresponding points of these snapshots 
can be seen in the light curve of Fig.~\ref{fig:flare750_lc}.

The plot of the density ($\rho$)
(top row in Fig.~\ref{fig:flare750_states}) shows a clear structure
in the upper funnel region that appears as the flare transitions from one
state to the next. This structure is also characterized by low magnetization and high
plasma~$\beta$. According to the definition given above, it
constitutes a clear current sheet which, in the bright-state, is
surrounded by regions with high values of magnetization ($\sigma$) (middle
row in Fig.~\ref{fig:flare750_states}) and low plasma~$\beta$ (bottom row in
Fig.~\ref{fig:flare750_states}). These are ideal conditions for radiating 
non-thermal radiation. This current sheet in the bright state can
be characterized as a plasmoid chain, similar to those seen in
localized PIC simulations (\citealt{ball2018slope}; \citealt{sironi2014relativistic}; \citealt{petropoulou2018steady}).

It is important to emphasize that this increase in magnetization $\sigma$ and decrease in plasma 
~$\beta$ in the bright state of the flares (conditions ideal for the activation of flares) are also 
found at the boundaries of the disk. These are possible regions for flare generation through high magnetic turbulence in the plasma, as mentioned in subsection~\ref{sec:radiation}. Fig.~\ref{fig:flare750_states} 
clearly shows that the 
creation of the current sheet inside the funnel region is a phenomenon perfectly connected to the 
organization of the magnetic field and results in the creation of conditions (magnetization and plasma 
~$\beta$) which will activate non-thermal particles in the current sheet itself and along the disk boundary.

Indeed, in accordance with the above, the areas that radiate are the current sheet and the disk 
boundary as shown in Fig.~\ref{fig:flux_slope_efficiency_data750}. 
Fig.~\ref{fig:flux_slope_efficiency_data750} illustrates the radiation flux (in mJy), the slope $p$ and the efficiency $\epsilon$
at time $t=910$~min (bright state of Flare B that we investigate). The visible boxes in each plot correspond to the identification of current sheets and their environment as shown in Fig.~\ref{fig:env_cs}. Within each box, all  plasma quantities are averaged in order to characterize this specific reconnection region. However, only the plasma within the current sheet layer - determined by the constraints and limits discussed in Section~2.3 - is considered to have the specific slope and efficiency,  producing the reported flux.
Two-thirds of the non-thermal 
radiation flux comes from the disc boundary, where the turbulent plasma dominates, while  one-third comes from the current sheet.

Similar phenomena can be observed in  Flare A at time $t=136$~min. A clear current sheet has produced a plasmoid chain, and the local environment is characterised by high magnetization and low plasma $\beta$ (see Fig.~\ref{fig:rho_magn_plasmab_data443}). Accordingly, the slope $p$ and the efficiency $\epsilon$ 
(middle and right panel of Fig.~\ref{fig:flux_slope_efficiency_data443}) peak at the current sheet and produce the radiation flux (left panel of Fig.~\ref{fig:flux_slope_efficiency_data443}), which strengthens our 
conclusions about how the flares are generated in this particular calculation.

\section{Conclusions}
We analyzed data from the GRMHD simulation of \citet{nathanail2020}, specifically 
focusing on model D, which is characterized as a SANE multi-loop model. Unlike MAD simulations, this model describes the accretion disk of a black hole  without the production of a stable jet.

By applying a thermal radiation model to this particular simulation enabled us
to reproduce the lightcurve at 230~GHz, from which we were able to
calibrate the simulation variables. 
Encouragingly, the model's variability demonstrated good agreement with observations, prompting us to conduct larger simulations with extended time evolution.

In order to calculate the non-thermal radiation from the simulation, we 
introduced a novel method that identifies current sheets and places with high magnetic turbulence. This identification based on the magnetic field polarity reversals, is done by 
constraining  primarily the current density together with the micro-physical plasma parameters, like magnetization and plasma~$\beta$. After the identification is done, all quantities are averaged in the local environment.
 The constraints we placed on the parameters provided clear spatial
boundaries within which to apply the model that calculates the non-thermal
radiation. 
This association not only advances our comprehension of the fundamental
processes at play, but also provides a crucial framework for future
investigations in similar contexts. 

The results for the lightcurve at $2.2~\mu$m are very encouraging as they
produce not only small flares up to 2~mJy but also a couple of brighter ones that
reach 7~mJy. The duration of such flares is also consistent with
observations. Further analysis of the flare spectral index demonstrated the
success of the model in reproducing the observations. It is interesting that during the evolution of the flare the spectrum follows a
power law similar to those given by the observations with spectral index $a=-0.44$.

In summary, we emphasize once again that a very important result of this work is the identification of the
source of flaring non-thermal radiation in GRMHD simulations. We have clearly shown that
such flares most probably originate in current sheets and their associated
plasmoid chains in the funnel area, as well as in the disk boundary due to magnetic turbulence. 

\section*{Acknowledgements}

ID is supported by the Hellenic Foundation for
Research and Innovation (HFRI) under the 4th Call for HFRI PhD
Fellowships (Fellowship Number: 9239).
Support comes from the ERC Advanced Grant ``JETSET:
Launching, propagation and emission of relativistic jets from binary
mergers and across mass scales'' (Grant No. 884631).
CMF is supported by the DFG research grant ``Jet physics on horizon scales and beyond" (Grant No.  443220636) within the DFG research unit ``Relativistic Jets in Active Galaxies" (FOR 5195).
This work was supported by computational time granted from the National Infrastructures for Research and Technology S.A. (GRNET S.A.) in the National HPC facility - ARIS - . Simulations
were performed also on the GOETHE-HLR cluster
at CSC in Frankfurt

\section*{Data Availability}
The data underlying this article will be shared on reasonable re-
quest to the corresponding author.

\bibliographystyle{aa}
\bibliography{Literature.bib}

\end{document}